%% file: main.tex
\documentclass[onecolumn,,10pt]{IEEEtran}

\usepackage{enumitem}
\input{preamble.tex}

\usepgfplotslibrary{colorbrewer}

\usepackage{textcomp}
\usepackage{colortbl}
\usepackage{subfigure}
\usepackage{array}
\usepackage{courier}
\usepackage{wrapfig}
\usepackage{pifont}
\usetikzlibrary{chains,backgrounds}
\usetikzlibrary{intersections}
\usetikzlibrary{pgfplots.groupplots}
\usepgfplotslibrary{fillbetween}
\usetikzlibrary{arrows.meta}
\usepackage{pgfplotstable}
\usepackage[super,compress,sort,comma]{natbib}
\usepackage{setspace}
\usetikzlibrary{math}
\usetikzlibrary{matrix}
\usepackage{xstring}
\usepackage{xspace}
\usepackage{flushend}
\makeatletter
\renewcommand\section{\@startsection {section}{1}{\z@}%
  {-2ex \@plus -1ex \@minus -.2ex}%
  {1ex \@plus.1ex}%
  {\Large\bfseries\scshape}}
\renewcommand\subsection{\@startsection {section}{1}{\z@}%
  {-1ex \@plus -.25ex \@minus -.2ex}%
  {0.1ex \@plus.0ex}%
  {\fontsize{11}{10}\selectfont\bfseries\sffamily\color{DodgerBlue4}}}
\renewcommand\subsubsection{\@startsection {section}{1}{\z@}%
  {0ex \@plus -.5ex \@minus -.2ex}%
  {0.0ex \@plus.5ex}%
  {\fontsize{9}{9}\selectfont\bfseries\sffamily\color{Red4}}}
\renewcommand\paragraph{\@startsection {section}{1}{\z@}%
  {-1.5ex \@plus -.5ex \@minus -.2ex}%
  {0.0ex \@plus.5ex}%
  {\fontsize{9}{9}\selectfont\itshape\sffamily\color{teal!50!black}}}

\makeatother  
\makeatletter
\pgfdeclareradialshading[tikz@ball]{ball}{\pgfqpoint{-10bp}{10bp}}{%
  color(0bp)=(tikz@ball!30!white);
  color(9bp)=(tikz@ball!75!white);
  color(18bp)=(tikz@ball!90!black);
  color(25bp)=(tikz@ball!70!black);
  color(50bp)=(black)}
\makeatother

\renewcommand{\captionN}[1]{\caption{\color{CadetBlue4!80!black} \sffamily \fontsize{9}{11}\selectfont #1  }}
\tikzexternaldisable 
\newif\iftikzX

\newif\ifFIGS
\FIGSfalse  
\FIGStrue
\title{ \sffamily \fontsize{20}{24}\selectfont Long-range Event-level Prediction and Response Simulation for Urban Crime and Global Terrorism with Granger Networks
} 

\author{\sffamily  \fontsize{10}{12}\selectfont  Timmy Li$^{1,2,4}$, Yi Huang$^{1,2}$,  James Evans$^{3,5}$ and Ishanu Chattopadhyay,$^{1,2\bigstar}$\\ 
\vspace{10pt}

\sffamily  \fontsize{10}{12}\selectfont
$^{1}$Institute of Genomics and Systems Biology,\\
$^{2}$Department of Medicine,\\ 
$^{3}$Department of Sociology, \\ and $^{4}$Department of Computer Science\\University of Chicago, Chicago, IL, 60637, USA\\
$^{5}$Santa Fe Institute, Santa Fe NM 87501, USA 
\vskip 1em
$^\bigstar$To whom correspondence should be addressed: e-mail:  \texttt{ishanu@uchicago.edu}.}
\begin{document}
\maketitle
\vspace{-15pt}

\begin{abstract} \bf \sffamily \fontsize{10}{12}\selectfont \noindent
Large-scale trends in urban crime and global terrorism are well-predicted by socio-economic drivers~\cite{ferdinand70,Cohen1979588,cohen81}, but focused, event-level predictions have had limited success~\cite{bowers04,Chainey2008,fielding12,mohler11,mohler15}. Standard machine learning approaches are promising~\cite{pmid28437486}, but lack interpretability, are generally interpolative, and ineffective for precise future  interventions with costly and wasteful false positives. Such attempts  have neither adequately connected with  social theory, nor analyzed  disparities between urban crime and differentially motivated acts of societal violence such as terrorism. Thus, robust event-level predictability  is still suspect, and  policy optimization via simulated interventions remains unexplored. Here, we are introducing Granger Network inference as  a new forecasting  approach for individual  infractions with demonstrated performance far surpassing past results, yet transparent enough to validate and extend social theory.   Considering the problem of predicting crime in the City of Chicago,  we  achieve an average AUC of $\approx 90\%$ for events predicted a week in advance within spatial tiles approximately $1000$ ft across. Instead of  pre-supposing that crimes   unfold  across contiguous spaces akin  to  diffusive  systems~\cite{mohler11,mohler15}, we learn  the local   transport rules from data. As our key insights, we uncover indications of  suburban bias~\cite{lipton1977poor,meyer16,jackson1987crabgrass,duany2001suburban,logan2002suburban,lazare2001america,young2002inclusion} | how law-enforcement response is modulated  by socio-economic contexts with disproportionately  negative impacts  in the inner city  | and how the  dynamics of  violent and property crimes co-evolve and constrain each other | lending  quantitative support to controversial pro-active policing policies~\cite{kelling96,bratton2009turnaround,mess07,harcourt06}. To demonstrate broad applicability to  spatio-temporal phenomena, we analyze  terror attacks in the middle-east in the recent past, and achieve an AUC of $\approx$ 80\% for predictions made a week in advance, and within spatial tiles measuring approximately 120 miles across. We conclude  that  while  crime operates near an equilibrium quickly dissipating perturbations, terrorism does not. Indeed terrorism aims to destabilize social order, as shown  by  its dynamics being susceptible  to run-away  increases in event rates  under small perturbations. 
  \end{abstract}

  \section*{Introduction}
  Crime and criminality have  been  undeniable aspects of the human condition since inception, as evidenced by  recorded history going back millennia~\cite{hammu04}.
  Over time, the rise of cities and the development of the urban social space created unique opportunities for crime~\cite{doi:10.1111/j.1745-9125.1976.tb00027.x,NBERw5430,shichor79,wirth38}, resulting in new challenges for prevention and policing.
 The recent emergence of ubiquitous   data driven modeling has sparked interest in \textit{predictive policing}: the possibility of predicting  crime, before it happens.
 In this study, we conceptualize this problem as that of modeling and prediction of a system of spatio-temporal point processes unfolding in the social context. We report a fundamentally  new approach to predict urban crime in space and time at the level of individual events, with predictive accuracy far greater than what has ever being achieved in the past. Beyond predicting the when and the where of the next infraction,  our new tools allow us to probe for enforcement biases, and garner deep insight into the nature of the dynamical processes that drive criminality in urban spaces.
 With an analysis framework applicable to general spatio-temporal phenomena, we  compare and contrast urban crime with terrorism, and find important similarities and distinctions. Our results indicate that while the dynamics of crime appears to operate in or around a stable equilibrium, terrorism is essentially an unstable system of interdependent events where small perturbations might be catastrophic.

Successful efforts to identify and explain  trends in urban crime go back at least half a century~\cite{ferdinand70,Cohen1979588,cohen81}.
Classical investigations into the  mechanics of  criminality  gave way to the possibility of event-level prediction of criminal infractions, in a manner that would make it possible to preemptively intervene, and ultimately engineer  urban spaces with lower  crime rates.
These efforts have reported  multi-variate modeling of time-invariant spatial distribution  of  hotspots~\cite{Wang2012,liu03,caplan17}, and have also included time-varying attributes~\cite{bowers04,Chainey2008,fielding12} to estimate both long and short term dynamic risks. A particularly visible approach to predictive policing is based on
the use of epidemic-type aftershock sequences (ETAS)~\cite{mohler11,mohler15} originally developed to model seismic phenomena. More recently, the application of standard deep learning architectures~\cite{pmid28437486} have been reported. While these approaches underscored  the potential of predictive policing, many have limited out-of-sample  performance~\cite{mohler11,mohler15}. Machine learning strategies~\cite{pmid28437486} have arguably  performed better,
but performance results generated with the common off-the-shelf tools ranging from random forests to neural nets have at least one important caveat: the common  approach of deleting a random sample of events |  training a machine learning system on the remaining data | and finally validating on the deleted sample  |  is  interpolative. Claims that good prediction on such interpolative validation schemes automatically translate to good event predictions in   the actual out-of-sample future time periods is, at best, strongly suspect. Additionally, most machine learning systems are black boxes with little or no insight into the sociology of the underlying phenomena | we learn nothing about the system, its rules of organization or have insight into how and where we can possibly intervene to modulate the course of its evolution.

Here, we show that urban crime may be predicted   precisely, reliably, and early enough that direct local intervention |  as well as high level  policy optimization from accurately predicted field impact |  becomes a practical strategy. We learn for a certain number of years (3) from recorded event logs, and then validate on events in the following year beyond those in the training sample. Using incidence data from the City of Chicago, our new spatio-temporal network inference algorithm, seeks out past patterns of event occurrences, and uses these inferred patterns or rules to construct a communicating network (the Granger Net) of local estimators, to ultimately predict future infractions. 

We make predictions 1) separately for violent and property crimes, 2) individually within spatial tiles roughly $1000 ft$ across, 3) approximately a week in advance, and 4) with AUCs ranging approximately between $80-99\%$ across the city.
Our approach outperforms past efforts significantly, on account of  realizing a predictive framework with little pre-defined structure on one hand, and vastly improving the sample complexity required for inference on the other  (See Discussion for comparison with competing approaches).

While not assuming  predefined constraints,  we also do not employ  off-the-shelf neural networks (NN) or other standard learning architectures. Indeed, unlike NNs which use  fixed non-linear activation functions to model influence propagation, we learn  the local transport rules from data. These local  rules are learned as finite state probabilistic transducer models, which significantly  outperform NNs in numerical experiments  in learning compact models of stochastic processes (See Supplementary text for explicit examples).

The  dependencies we infer are not constrained to  be local. In contrast  to what we expect in diffusive systems, social rules of interactions are  not required to mimic laws of physics; there is no guarantee that influence diffuses in any orderly fashion, or that events far off across the city will  have a weaker influence compared to those physically near in space or time. Thus, a computational approach that \textit{discovers}  the emergent social topology on which the dynamics of interest unfolds | as opposed to assuming either some rigid  model structure or a diffusive framework | is key to our achieved  performance.
 
With our precise predictive apparatus in place, we run a series of computational experiments that perturb the rates of violent and property crimes, and log the resulting alterations in future event rates  across the city. By inspecting the effect of SES variables on the perturbation response, we can then  investigate if enforcement and policy biases modulate outcomes. Our analysis suggests that the enforcement response to worsening crime rates is strongly modulated by the status of socio-economic variables, and disadvantaged neighborhoods might suffer from  resources being pulled away to wealthier counterparts.

Additionally, our analysis reveals direct evidence as to  how violent and property crimes  co-evolve and influence one another, indicating  that increase in one typically down-regulates the other. These  dependencies  potentially indicate that somewhat  controversial  pro-active enforcements, $e.g.$ the \textit{broken window policy}~\cite{kelling96}, might be justified in an operational sense. While this does not negate   objections  from the viewpoint of social theory~\cite{sri06}, or minimize the issues of incorrect, overzealous, and improperly incentivized police practices~\cite{childress2016}, our analysis  suggests that | if properly implemented |   addressing property and non-violent infractions might indeed have a significant suppressive effect on violent crime.

Further, we shed light on the continuing debate on the correct choice  of neighborhood boundaries in urban crime modeling~\cite{sherman89,wool2002,mears06,Weisburd2014}. We demonstrate that influence often is communicated over large distances, and decay slowly, and perhaps more importantly, the ``correct'' choice of spatial scale  is less of an issue in sophisticated learning algorithms, where the optimal scales can be inferred automatically.

Finally we ask how the dynamics of urban crime compares to  event patterns in other acts of societal violence, specifically terrorism. Using data from the Global Terrorism Database (GTD), and focusing on the relatively recent $5$ year time-frame  in the  Middle-east, we carry out predictive analysis similar to the urban crime case. We find that predictability is lower (average AUC $\approx 80\%$), but not by much, for the case of terrorism; the drop in performance is most likely is due to the lower frequency of recorded events. We find that event influences propagate rapidly and widely | similar to the  crime scenario. However, the nature of the dynamics, as revealed by our perturbation analyses,  is markedly different between these two categories of aberrant social behavior | crime operates near or around a dynamic equilibrium, where perturbations generally tend to dissipate quickly, whereas terrorism appears to operate far from equilibrium, and small perturbations can be  destabilizing and  rapidly increase event rates. 

 To the best of our knowledge, analysis of the  perturbation response of  data-driven models to probe  underlying social constructs has not been reported before. Even with the current  limitations, this new addition to the toolbox of computational sociology allows for  direct  validation of complex theory from observed event incidence, supplementing the use of subjective measurable proxies, and potential biases in questionnaire-based  data collection strategies. While such classical approaches have  unquestionably broadened our understanding of the societal forces  shaping the urban landscape, and inspired  social theory to investigate  the nature,  correlates, and causal drivers of criminality~\cite{suth42,Sampson918,miethe91,braga14},  they  do not attempt to forecast individual  events.
In this study, we show that the ability to predict such events opens new doors to not only precise intervention possibilities, but to a whole new set of computational tools.

\ifFIGS
\begin{figure}[!ht]
  \tikzexternalenable

  \centering  
\iftikzX
  \input{Figures/fig0}
  \else \includegraphics[width=0.97\textwidth]{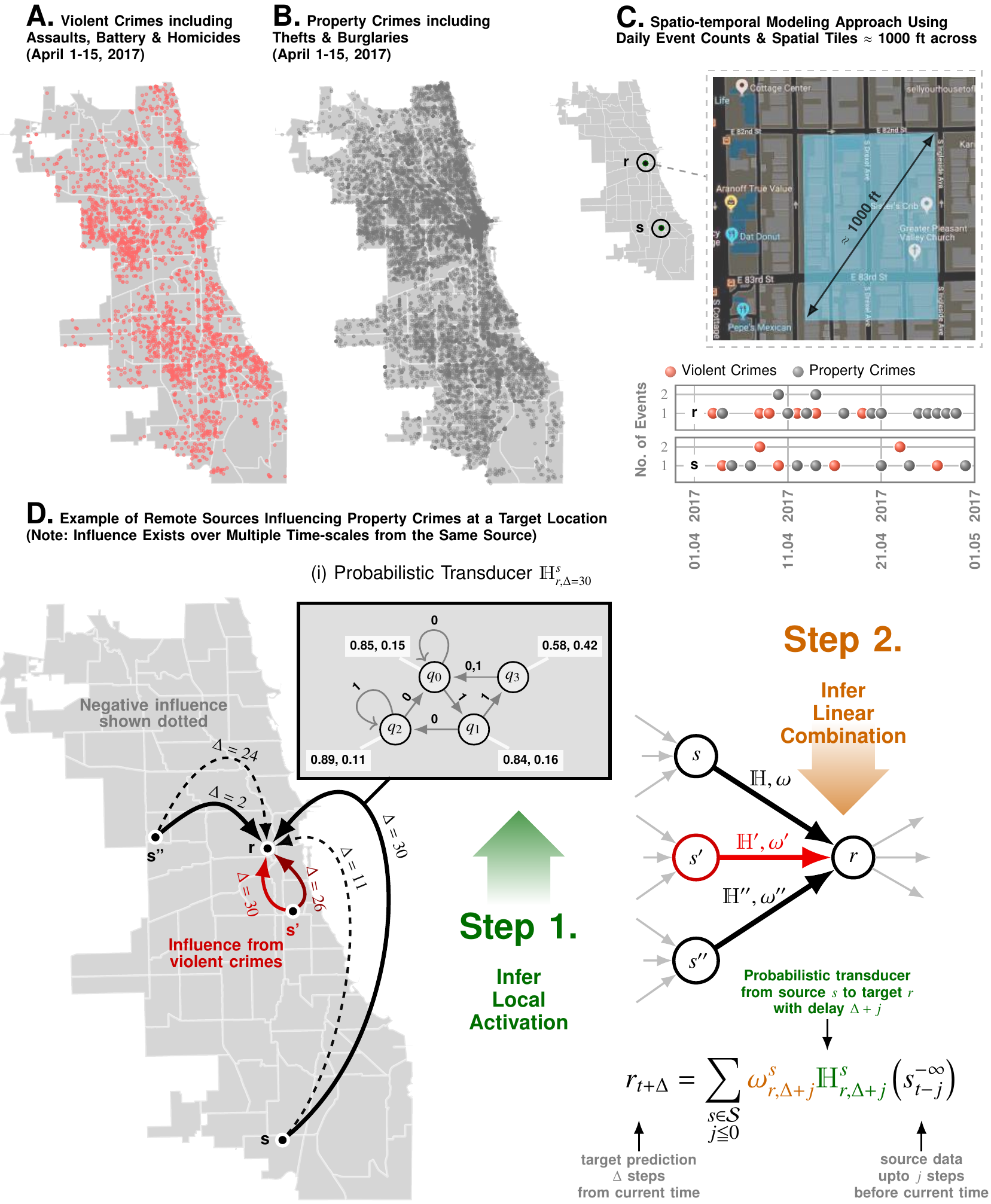}
  \fi

  \vspace{-1pt}

  \captionN{\textbf{Crime Data \& Modeling Approach}. Plates A and B show the recorded infractions within the 2 week period between April 1 and 15 in 2017. Place C illustrates our modeling approach: We break up the city into small spatial tiles that are about 1.5 times the size of an average city block, and compute models that capture multi-scale dependencies between the sequential event streams recorded at distinct tiles. In this paper, we treat violent and property crimes separately, and show that these categories have intriguing cross-dependencies.  Plate D illustrates our modeling approach. For example, to predict the property crimes at some spatial tile $r$, we proceed as follows: step 1) we infer the probabilistic transducers that estimate the event sequence at $r$ by using as input the sequences of recorded infractions (of different categories)   at potentially all remote locations ($s,s',s''$ shown), where this predictive influence might transpire over different time delays (a few shown on the edges between $s$ and $r$). Step 2) Combine these weak estimators linearly to minimize zero-one loss. The inferred transducers can be thought of as inferred local activation rules, which are then linearly composed, which reverses the approach of linearly combining input and then passing through fixed activation functions in standard neural net architectures. The connected network of nodes (variables), with  probabilistic transducers on the edges comprises the Granger Net.  }\label{fig0}
  \vspace{-15pt}
\end{figure}
\else
\refstepcounter{figure}\label{fig0}
\fi
\ifFIGS
\begin{figure*}[!ht]
  \tikzexternalenable

  \centering 
\iftikzX
 \hspace{-10pt} \input{Figures/figpred_}
\else \includegraphics[width=0.99\textwidth]{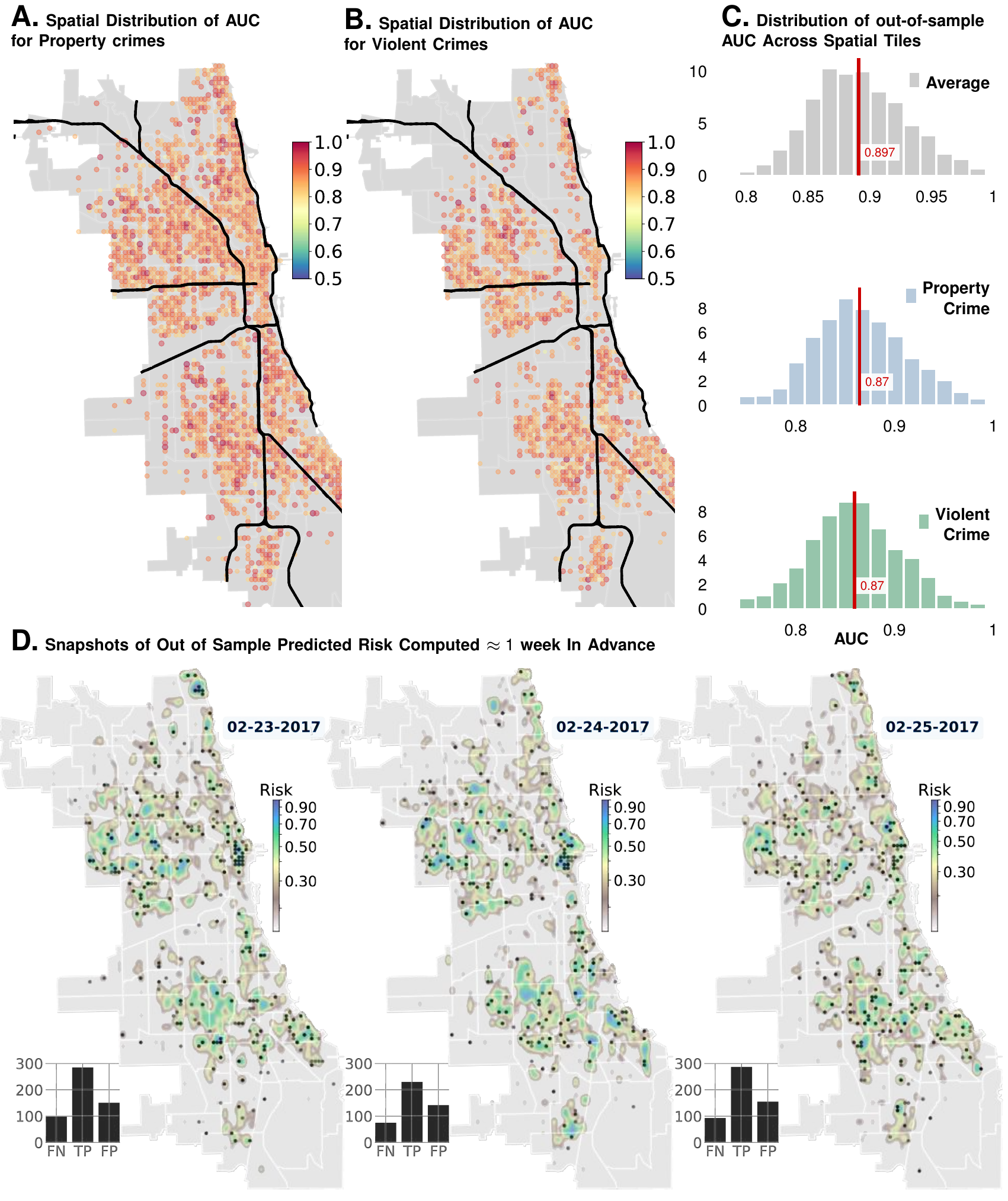}
  \fi

  \vspace{-5pt}

  \captionN{\textbf{Predictive Performance of Granger Nets. } Plates A an B illustrates the out-of-sample area under the receiver operating characteristics curve (AUC) for predicting violent and property crimes respectively. The prediction is made a week in advance, and the event is registered as a successful prediction if we get a hit within $\pm 1$ day of the predicted date. Plate C illustrates the distribution of AUC on average, and individually for violent and property crimes. Our mean AUC is close to $90\%$. Plate D illustrates the risk computed $7$ days in advance for 3 consecutive days in 2017 February (out-of-sample). The red dots are actual events (violent or property crimes), and the computed risk is shown as an overlay. Event predictions at individual spatial tiles is used to construct the continuous risk intensity  map by summing Gaussian densities  centered at each predicted event location. The variance of the Gaussian densities is tuned (in the course of training) to maximize recall. The risk shown is normalized within each day.  }\label{fig1}
\end{figure*}
\else
\refstepcounter{figure}\label{fig1}
\fi
\ifFIGS
\begin{figure}[t]
  \tikzexternalenable

  \centering 

 \iftikzX
 \input{Figures/figbias}
 \else
\includegraphics[width=0.99\textwidth]{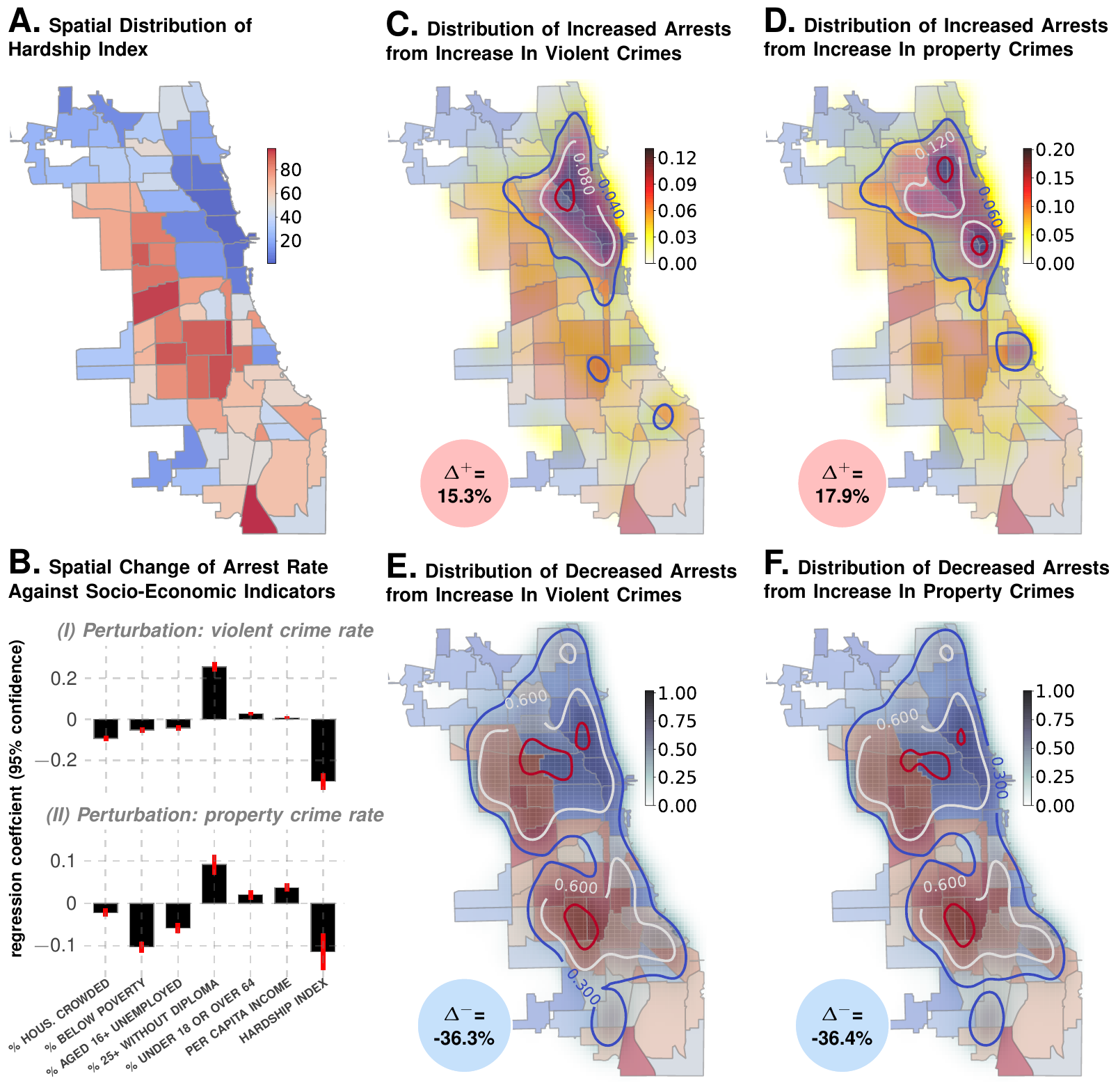}
  \fi

  \vspace{-10pt}

  \captionN{\textbf{Estimating Bias. } Plate A illustrates the distribution of hardship index. Plates C, D, E, and F show the biased response to perturbations in the crime rates.
    With a 10\% increase in each of violent and property crime rates, we  have approximately a $30\%$ decrease in arrests when averaged over the city. However, the spatial distribution of locations which experience a positive vs a negative change in the rate of arrests reveals a strong preference for favorable socio-economic indicators for the former. Thus, if neighborhoods are doing better socio-economically, increasing crime seems to predict increased arrests as expected. A strong converse trend is observed in our predictions for neighborhoods doing worse, suggesting that under stress, the wealthier neighborhoods drain resources from their disadvantaged counterparts. Plate B illustrates this more directly via a multi-variable regression, where the hardship index is seen to have  a strong negative contribution. }\label{fig2}
\end{figure}
\else
\refstepcounter{figure}\label{fig2}
\fi
\ifFIGS
\begin{figure}[!ht]
  \tikzexternalenable

  \centering 

 \iftikzX
 \input{Figures/figspace}
 \else
\includegraphics[width=0.99\textwidth]{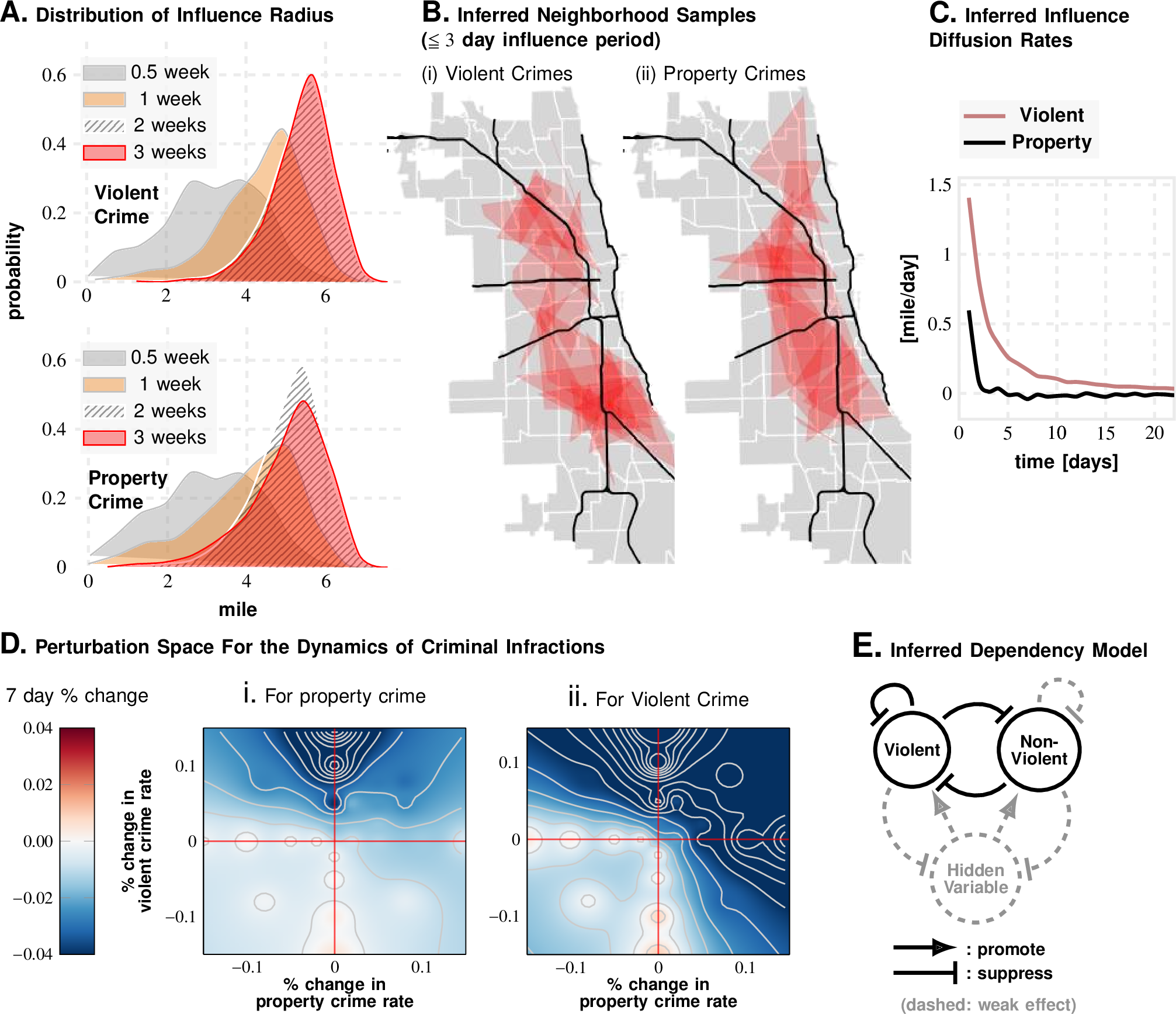}
  \fi

  \vspace{0pt}

  \captionN{\textbf{Influence Diffusion \& Perturbation Space.} If we are able to infer a  model that is usefully predicts the  event dynamics at a specific  spatial tile (the target) using observations at a source tile $\Delta$ days in future, then we say that the source tile is within the influencing neighborhood for the target location with a delay of $\Delta$.   Plate A
illustrates the distribution of spatial radius of influence for 0.5, 1, 2 and 3 weeks of time, for violent (upper panel) and property crimes (lower panel). We note that the influence neighborhoods, as defined by us,  are large, and tend to approach a radius of $6$ miles eventually. Given the geometry of the City of Chicago, this maps to  a substantial percentage of the total area of the urban space under consideration, showing that crime here has demonstrable long-range and almost city-wide influence on average. Plate B illustrates the extent of a few of the inferred  neighborhoods at a time delay of at most $3$ days, which as noted, are not limited to local city blocks. Plate C illustrates the average rate of influence diffusion measured, as before, by the number of predictive models inferred that transduce influence as we consider longer and longer time delays. We note that the rate of influence diffusion falls rapidly; for property crimes, the rate goes to zero in about a week, whereas for violent crimes, the influence keeps diffusing even after three weeks.
Plate D illustrates the multi-dimensional perturbation space constructed from probing the inferred Granger Net with $1-10\%$ perturbations in violent and property crime rates. We see that violent and property crimes are anti-thetical: increase in one leads to suppression in the other. Most importantly, the effect of suppressing violent crimes by increasing property crimes (moving right on the x-axis in plate D(i)), suggests important policy implications. Plate E summarizes the key dependency patterns: notably violent crimes are more strongly self-limiting. The fact that we have all suppressive relationships suggests a hidden variable that keeps the dynamics alive. Compare the dependency pattern in terror events, where we could infer no such  suppressive relation. }\label{fig3}
\end{figure}
\else
\refstepcounter{figure}\label{fig3}
\fi

\ifFIGS
\begin{figure}[!ht]
  \tikzexternalenable

  \centering 

 \iftikzX
 \input{Figures/figspace_terror}
 \else
\includegraphics[width=0.99\textwidth]{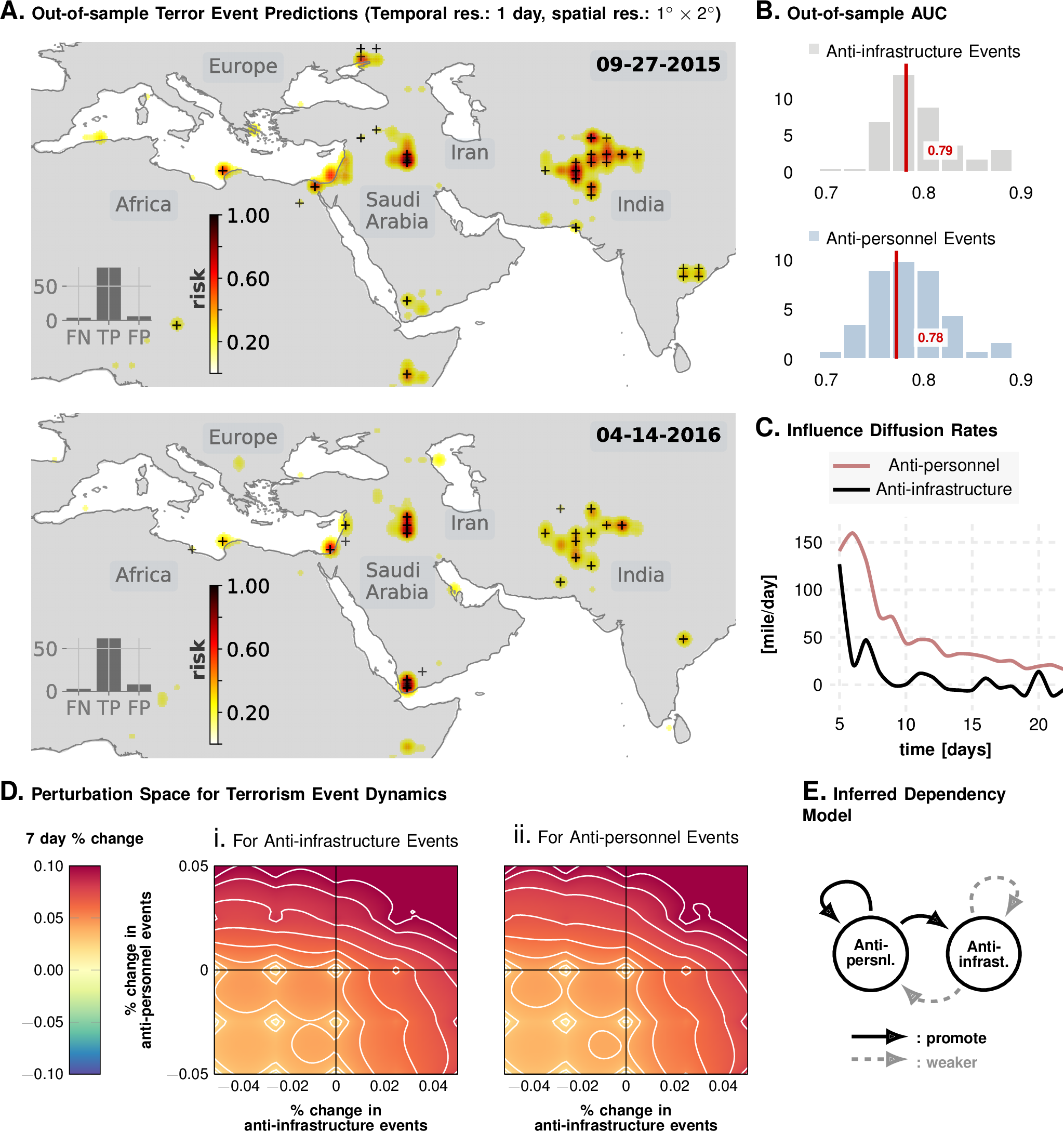}
  \fi

  \vspace{-5pt}

  \captionN{\textbf{Terror Event Prediction \& Perturbation Space.} Plate A shows out of sample predictions for terror events in the middle east and surrounding regions for two specific days in 2015 and 2016 respectively. The overall distribution of AUC for anti-infrastructure events and anti-personnel events is shown in plate B. In Plate C,  we illustrate the influence diffusion rates for the two classes of terror events (Fig.~\ref{fig3}, Plate C). 
We see that the diffusion rates decay rapidly as before, and the rate for  anti-infrastructure events reduces to zero in about $7--10$ days, whereas the rate for anti-personnel events keeps diffusing even after three weeks. 
Plate D illustrates the multi-dimensional perturbation space constructed, as before, by probing the inferred Granger Net with $1-10\%$ perturbations in the two classes of events. Unlike the dynamics of crime,  terrorism does nit seem to have any self-limiting behavior: injecting positive perturbations in either kind of event stream strongly increases the future probability of both categories of events.  Plate E summarizes the key dependency patterns: notably we find no suppressive dependencies, implying a very distinct dynamical pattern compared to urban crime.  }\label{fig4}
\end{figure}
\else
\refstepcounter{figure}\label{fig4}
\fi
 \section*{Method \& Materials}
 \subsection*{Data Source}
 Event incidence data  for this study is obtained  from the City of Chicago Data Portal~\cite{w2019}.  The log includes spatio-temporal event localization along with the nature, category,  and a brief description of the recorded incident.  Additional  information on the number of arrests made during each event is also included.  The log  is updated daily, keeping current with a lag of $7$ days.
We only use data between 2014-2017 ($3$ years for model inference, and last $1$ year for out-of-sample validation) for the prediction results shown in Figure~\ref{fig0}. The evolving nature of the urban scenescape~\cite{silver2016scenescapes} necessitates that we restrict the modeling window  to a few years at a time. The length of this window is decided by trading off the loss of performance from shorter data streams to that from the  evolution of the underlying generative processes for wider windows (See Supplementary text). In this study, we consider   two broad categories of criminal infractions: \textit{violent crimes} consisting of homicides, assault, and battery, and \textit{property crimes} consisting of  burglary, theft and motor vehicle thefts. The number of individuals arrested  during each recorded event is considered as a separate variable to be modeled and predicted, which allows us  to investigate the possibility of  enforcement biases in subsequent perturbation analyses.
 
Additionally, we  use data on socio-economic variables available at the portal corresponding to the Chicago community areas and census tracts, including the  $\%$ of population living in crowded housing, those residing below poverty line, those unemployed at various age groups, per capita income, and the urban hardship index~\cite{nathan89}.

To contrast urban crime against a motivationally  different  phenomenon of societal violence, we analyzed data from 2012 to 2016 from the open-source Global Terrorism Database (GTD), with the last year used as the out-of-sample validation set. To ensure  similarity  to the urban crime scenario, we chose to consider events in two categories: 1) anti-infrastructure events including bombing, explosions, and facility and infrastructure attacks, and 2) anti-personnel  events including armed assault, hostage-taking, barricade incident, hijacking, assassination and kidnapping. We also consider the number of casualties as a third variable to be modeled, analogous to the \textit{number of arrests} in the crime scenario.
 \subsection*{Event Log Processing: Spatial and Temporal Discretization \& Event Quantization}
 The  event log  is processed  to obtain  time-series of events of interest, stratified by occurrence locations. This is accomplished by choosing a spatial discretization, and focusing on an individual spatial tile at a time, which allows us to represent the event log as a collection of sequential event streams (See Fig.~\ref{fig0}, plate C). Additionally, we discretize time, and consider the sum total of events recorded within each time window. 

 Coarseness of these  discretizations reflects  a trade-off between computational complexity, and event localization  in space and time. The spatial and the  temporal discretizations  are not independently chosen; a finer spatial discretization dictates  a coarser temporal quantization, and vice versa to prevent either long no-event stretches, or long  periods of contiguous event records, both of which are detrimental to obtaining reliable  predictors. In this study, we fixed the temporal quantization to $1$ day, and chose a spatial quantization such that we have high empirical entropy rates on average for  the set of time series obtained (See Supplementary text for details).
This resulted in spatial tiles measuring $0.00276\textrm{\textdegree} \times 0.0035\textrm{\textdegree}$ in latitude and longitude respectively, which is approximately $1000'$  across, roughly corresponding to the area of under $4$ city blocks. Note any two points within our spatial tile are at most in  neighboring blocks.  We  dropped from our analysis the tiles that have too low a crime rate ($< 5\%$ of days within the modeling window  had any event recorded) to reduce computational complexity, ending up with a total of $N=2205$ spatial tiles.

Thus, we  end up with three different integer-valued time series at each spatial tile: 1) violent crime ($v$), 2) property crime ($u$) and 3) number of arrests ($w$). We ignore the magnitude of the observations, and treat them as Boolean variables; thus our models simply predict the presence or absence of a particular event type within a particular spatial tile (within couple of blocks), and within a particular observation window ($1$ day).
\subsection*{Inferring Stochastic Generators of Spatio-temporal Cross-dependence}
\def\S{\mathcal{S}}
\def\L{\mathcal{L}}
\def\E{\mathcal{E}}
\def\H{\mathds{G}}
\def\D{\mathcal{D}}
\def\H{\mathds{H}}
Let $\L=\{\ell_1,\cdots,\ell_N\}$ be the set of spatial tiles, and $\E=\{u,v,w\}$ be the set of event categories as described in the last section.  At location $\ell \in \L$ for variable $e \in \E$, at time $t$, we have  $(\ell,e)_t\in \{0,1\}$, with  $1$ indicating the presence of at least one event. The set of all such combined variables (space + event type) is denoted as $\S$, $i.e.$, $\S = \L \times \E$. Let $T=\{0,\cdots,{M-1}\}$ denote the training  period consisting of $M$ time steps. Since for any time $t$, $(\ell,e)_{t}$ is a random variable, our goal here  is to learn its dependency relationships with its own past, and with other variables in $S$, to  accurately estimate its future  distribution  for  $t>T$. 

To infer the structure of the Granger Net, we learn a finite state probabilistic transducer~\cite{ixc14,Mohri2004}  for each possible source-target  pair $s,r \in \S$. Given a sequence of events at the source, these inferred transducers estimate the distribution of  events at the target $r$ at some future point in time. Ability to estimate such  a non-trivial distribution indicates the presence of causal influence.
Here we assume that causal influence from the source to the target  manifests as the source being able to predict events occurring at the target, better than the target can do by itself; this interpretation follows from  Granger's~\cite{Granger1980} eponymous approach to statistical causality. Such influence is not restricted to be instantaneous; the source events might impact the target  with a time delay, $i.e.$, a specific model between the source and the target might predict events delayed by an a priori determined  number of steps  $\Delta_{max} \geqq \Delta\geqq 0$ specific to the model. Here, we model the influence structure for each integer-valued delay separately; thus for source $s$ and target $t$, we can have $\Delta_{max}+1$ transducers each modeling the influence for a specific  delay  in $[0,\Delta_{max})$. The maximum number of steps in time delay $\Delta_{max}$ is chosen a priori, based on the problem at hand.

While these influences or dependencies may be different for different delays, they also do not need to be symmetric between the source-target pairs. The complete set, comprising at most $\vert \S \vert^2 (\Delta_{max}+1)$ models, represents a predictive framework  for asymmetric multi scale spatio-temporal phenomena. Note that the number of possible models increase quickly, $e.g.$, for $\Delta_{max}=60$, with $2205$ spatial tiles, and three event categories,  the number of inferred models is bounded above by $\approx 2.6$ billion.

In this study we learn transducers  as crossed finite state probabilistic automata (XPFSA) models (See Supplementary text). 
Our approach consists of two key steps (See Fig.~\ref{fig0}, plate D): First, we infer  XPFSA models for all source-target pairs and all delays upto $\Delta_{max}$, and then in the second step, we learn  a linear combination of these transducers to maximize predictive performance.

\textit{Step 1:} Denoting the observed event sequence in the time interval $(\infty, t]$ at source $s$ as $s^{-\infty}_t$, the XPFSA $\H^s_{r,k}$ estimates  the distribution of events for target  $r$ at time step $t+k$. This is accomplished by learning an equivalence relation on the historical event sequences observed at the source $s$, such that equivalent histories induce an identical (or a nearly identical) future event distribution at the target $r$, $k$ steps in the future. Thus, for example, the XPFSA shown in Fig.~\ref{fig0} plate D, has four states, indicating that there are $4$ such equivalence classes of observations that induce the distinct output probabilities shown from each state.
Often this estimate is  not very precise due to the possibility of multi-scale and multi-source influence, $e.g.$, when the target $r$ is influenced by multiple sources, and with different time delays.

\textit{Step 2:} We employ a standard gradient boosting regressor for each target, to optimize the linear combination of
inferred transducers, and learn the scalar weights $\omega^s_{r,k}$ for source $s$, target $r$ and delay $k$,  such that the estimate:
\cgather{
  r_{t+\Delta} = \sum_{\substack{j\leqq 0\\s \in \S}} \omega^s_{r,\Delta+j} \mathds{H}^s_{r,\Delta+j}\left (s^{-\infty}_{t-j}   \right ) \label{eqF} }
\noindent{minimizes} the expected  loss. Note that the left hand side of Eq.~\ref{eqF} is a random variable, $i.e$, we are estimating a stochastic process indexed by $t+\Delta$, as a function of the observed sample path $s^{-\infty}_t$. Here, $\Delta$ is a pre-specified constant, which specifies how far into future we are making the prediction.

To compare with a standard neural net architecture, these probabilistic transducers may be viewed as local non-linear activation functions. Thus, while with neural networks we repeatedly compute affine combination of inputs  and apply fixed non-linear activation to the combined input and finally optimize the affine combination weights via backpropagation, here we first learn the local non-linear activations, and then optimize the linear or affine combination of the weak estimators. Optimizing the weights is significantly simpler | and a local operation | for us, and may be done with any standard regressor, even with a local NN. In contrast to recurrent neural nets (rnn), the role of the hidden layer neurons is partially taken over in our case by the states of the XPFSA, which are a priori undetermined both with regards to their multiplicity and their transition connectivity structure. Even with the significantly simplified computation, our approach is provably PAC-efficient~\cite{Valiant84}, $i.e.$, we can learn good models with high probability with relatively small sampling complexity.

  The non-trivial structure of the Granger Net emerges from not all models being useful; we do not ultimately use all $2.6$ billion models for crime prediction. We estimate the usefulness of a particular model by computing it \textit{coefficient of causality}~\cite{ixc14} ($\gamma$, see Supplementary text), which estimates the relative reduction in entropy in the predicted outcomes   over the entropy of the time-averaged target distribution, $i.e.$ the prediction we will get with no model inference. Throwing away these poor models reveals the complex predictive wiring of the system at hand, which in itself holds important clues to its dynamical characteristics.
  \subsection*{Computational \& Model Complexity}
  We assume the maximum time delay in the influence propagation to be  $60$ days, resulting in at most  $2,669,251,725$ inferred models, of which $61,650,000$ are useful  with $\gamma \leqq 0.01$. Model inference consumed approximately $200K$ core-hours on $28$ core Intel Broadwell processors, carried  out with incidence data over the period Jan 1, 2014 to December 31, 2016. Data from January 1 2017 to December 31, 2017 is used for out-of-sample validation. 
  \subsection*{Perturbation Analysis}
  Determination of stability characteristics is a central question in any system modeling. For the dynamics of criminal infractions, the question of stability has important sociological implications: how close is the system to a run-away behavior, where we experience increasingly large upticks in the event rates? Or is there some tangible evidence in the data at hand that  such instabilities are not likely in the foreseeable future?

  To answer these questions, we investigated the response of the system to bounded perturbations. The perturbations are injected  by  modifying the observed event sequences as follows:  to introduce a positive perturbation to the crime rates, we randomly replaced $0$s in the binary event streams with a sample from a Bernoulli($\theta$) distribution, where $\theta$  is chosen to reflect the desired increase in the event rate (See Supplementary text). To introduce a negative perturbation, $i.e.$, a reduction in the crime rates, we replaced $1$s in the discretized binary data streams with a sample from a Bernoulli($1-\theta$) distribution.
  
  We  experimented with  positive and negative  perturbations to both violent and property crime rates ranging between $1$ to $10\%$ of the observed rates. Response to the perturbed crime rates was measured as  the relative change  from the nominal baseline in the estimated time-average in the predicted event frequencies $1$ week into the future, corresponding to violent and property crimes, and the number of arrests.

  Results from the perturbation experiments shed light both on the stability characteristics of  crime in Chicago, and further allowed us to
  look for the evidence of biased enforcement response under stress.
 \section*{Results}
 Our key result is the  development of an efficient framework for
 event-level prediction of urban crime. For each spatial location, the inferred Granger Net  maps event histories to a raw risk score as a function of time | higher this value, higher the probability of an event of the target type occurring at that location, within the specified time window. However, to make crisp predictions, we must choose  a decision threshold for this raw score. Conceptually identical to the notion of Type 1 and Type 2 errors in classical statistical analyses, the choice of a threshold trades off false positives (Type 1 error) for false negatives (Type 2 error): choosing a small threshold  results in predicting a larger fraction of future events correctly, $i.e.$ have a high true positive rate (TPR), while simultaneously suffering from a higher false positive rate (FPR), and vice versa. The receiver operating characteristic curve (ROC) is the plot of the  FPR vs the TPR, as we vary this decision threshold. If our predictor is good, we will consistently achieve high TPR with small FPR resulting in a high area under the ROC curve denoted as the AUC; AUC measures  intrinsic performance, independent of the threshold choice. More importantly, the AUC is  immune to class imbalance (the fact that crimes are by and large rare events). An AUC of $50\%$ indicates that the predictor does no better  than random, and an AUC of $100\%$ implies that we can achieve perfect prediction of future events, with zero false positives.
 \subsection*{Predictability Achieved  In the City of Chicago}
 We can predict events approximately a  week in advance at the spatial resolution of $\approx 2 $ city blocks with a temporal resolution of $\pm 1$ day, with a false positive rate of less than $20\%$, with a median true positive rate of $78\%$. 
 Our prediction results are summarized in Fig.~\ref{fig1}, where plates A and B illustrate the geospatial scatter of AUC obtained for different spatial tiles, property and violent crimes respectively. Plate C shows the distribution of the AUC achieved where we predict a crime ignoring its category (top plate, mean AUC $\approx 90\%$), and where we predict property and violent crimes separately (middle and bottom plates, with mean AUC $87\%$ in both cases). These predictions are made $1$ week in advance, but we register a true hit if the predicted event transpires within $6-8$ days of the day the prediction is made.

 Plate D illustrates the predicted risk map for  $3$ consecutive days in 2017 February, overlaid with the locations of actual criminal infractions, with both violent and property crimes considered.
 Event predictions at individual spatial tiles is used to construct the continuous risk intensity  map by summing Gaussian densities  centered at each predicted event location. The variance of the Gaussian densities is tuned (in the course of training) to maximize recall (ratio of true positives to the sum of true positives and false negatives).
 The risk map shown in Fig.~\ref{fig1} Plate D is normalized within each day. Such normalized  maps can be directly used to optimize law enforcement deployments under constrained resources, where the recommendation will be to prioritize the peaks of the daily risk map.

 \subsection*{Enforcement \& Policy Bias: Resource Distribution Response Under Stress}
The results from the perturbation experiments suggest that under stress, well-off neighborhoods tend to drain resources disproportionately from disadvantaged locales. Our findings are summarized in Fig.~\ref{fig2}. We find that on small perturbations of the crime rate, the corresponding variation in law enforcement response, measured by variation in the number or predicted arrests, is very different in economically well-off neighborhoods. These neighborhoods see a roughly proportionate increase in arrest rate with increasing crime (as expected), whereas the arrest rate in the rest of the city crashes by a factor of nearly 3. This suggests that increased ``stress'' in the form of increased number of crimes causes the enforcement resources to be drained out of disadvantaged neighborhoods to support their better  socioeconomic counterparts. 

A multi-variable regression analysis (Fig.~\ref{fig2} plate B) also supports this conclusion. It shows that the change in arrest rate from perturbations that increase the violent and the property crime rates  have  a strongly negative contribution from hardship index, which is contradictory to what is expected in the absence of bias. Poorer neighborhoods have more crime, and thus, these socio-economic indicators should  contribute positively, if at all, to increase the arrest rate as response to increasing crime. The reverse association seen here is problematic, and potentially indicative of biases at the level of policy driving resource allocation in the city.

\subsection*{Temporal Memory \& Neighborhood Effects}
There is significant work in the literature aiming to understand the role of neighborhood organization in shaping urban interactions. We probed the topological structure emergent in the inferred dependencies to estimate the shape, size and organization of the neighborhoods that predict events at each location. The results, illustrated in Fig.~\ref{fig3} plates A and B, show that the situation is complex with the locally predictive neighborhoods varying widely in geometry and size.
Clearly, restricting our analysis to relatively small local communities within the city is less than optimal, and even attempting to determine a priori the correct scale of such organization that best predicts crime might not be even feasible. 

We then asked if the effect of criminal infractions  diffuse outward in space and time, and if the diffusion rate of influence can be meaningfully estimated. While the diffusion rates appear to  vary significantly from one location to the next, when averaged across the city, we  see a  rapid decay with time delay in the diffusion rates as shown in Fig.~\ref{fig3}, plate C. Note however that the diffusion rate for property crimes decays much faster compared to that from violent crimes. In particular, on average,  the diffusion rate from property crimes decays to zero in about $10$ days, while violent crimes stay relevant event after a couple of months.

\subsection*{Emergent Relationships Between Violent \& property crimes}

The system responses from our perturbation experiments is
used to estimate the multi-dimensional perturbation manifold, as illustrated in Fig.~\ref{fig3}, plate D. Plate D (i) shows the contours of the estimated time-averaged change in property crimes a week in future after the perturbations were introduced. Plate D(ii) illustrates the response for violent crimes. We can gain important insights into the underlying dynamical rules and constraints by imagining  likely system trajectories in this manifold, $e.g.$, by moving along the Y-axis in Plate D(i), $i.e.$ by increasing violent crimes, we suppress property crimes. And by moving along the X-axis in plate D(ii), $i.e.$, by increasing property crimes, we suppress violent crimes.

We observe that the two categories of  criminal infractions interact asymmetrically. As summarized in Fig.~\ref{fig3}, plate E, we see that these two categories  tend to counteract each other: increase in one leads to suppression in the other. Violent crimes seem to have significantly stronger self-limiting effect compared to property crimes. Most importantly, the effect of suppressing violent crimes by ``increasing'' property crimes (moving right on the x-axis in plate D(i)), suggests important policy implications (See Discussion).

\subsection*{Modeling  Terror Event Dynamics From GTD}
For the analysis of terror events, we use a spatial tiles that measure $1^\circ \times 2^\circ$ is latitudinal and longitudinal extent, and limited to the middle east (See Fig.~\ref{fig4}), this implies that our spatial tiles are roughly $\sqrt{69^2 +102^2} \approx 123.1$ miles along the diagonal. As with crime, we use a temporal quantization of $1$ day, and measure performance with prediction made a week in advance. We achieve AUCs close to $80\%$ ($78.8\%$ for anti-infra-structure events, and $77.9\%$ for anti-personnel events). The slightly reduced performance maybe attributed to the significantly lower rates of terror attacks compared to crime in Chicago, which in the later case allows us to model low  probability patterns better.

We also compute the diffusion rates for influence of  terror events, and it appears that anti-personnel events in this case is the one that decays slowly (See Fig.~\ref{fig4}, plate C, and compare with violent crimes in Fig.~\ref{fig3}, plate C). The influence of anti-infrastructure events dies down after about $10$ days, which is also the time that property crime influence take to approximately vanish, suggesting intriguing similarities between the two dynamical systems. In contrast, the perturbation analysis for terrorism brings forward important distinctions from urban crime.

\section*{Discussion}
Despite both violent and property crime in US  falling sharply over the past quarter century~\cite{gramlich2019}, major cities continue to have unacceptably high rates of violent and property crimes (See Fig.~\ref{fig0}, plates A and B). The distinctive  patterns  of urban crime have led sociologists to advance theories ranging from  the   connection  between urbanity  and immoral behavior, to the mechanics of  opportunity arising from denser population, more contact with the wealthy leading to  higher expected   pecuniary returns, and the demonstrably better chances of getting away with crime in cities~\cite{doi:10.1111/j.1745-9125.1976.tb00027.x,NBERw5430,shichor79,wirth38}. To theorize on the underlying causal factors shaping the  urban 
scene, scholars have suggested  mechanisms which breakdown informal social control~\cite{Shaw1942,Veysey1999156,Sampson1989774,Kubrin2003374}, and collective efficacy~\cite{Sampson918} to encourage and shape  criminal behavior.
In this study, we take a  computational approach: we  design a learning architecture  to reliably predict   individual crimes sufficiently before they happen, such that direct intervention becomes a possibility.

The contribution of this study is two fold: 1) our approach  significantly outperforms past  attempts to actionably predict crime, 2) we can use our inferred predictive structure to probe for fundamental insights into the underlying processes that drive urban crime.  The centerpiece of our approach is the Granger Net, which models complex  spatio-temporal dependencies inferred  without presupposing any particular model structure. Consequently,  we have a median TPR of  78\% with FPR not more  than $20\%$; in contrast to one of the prominent past efforts tuning pre-supposed ETAS-based models~\cite{mohler15} achieve a  $\leqq 10\%$ true positive rate  with no corresponding figure on false positives. While the lack of prior assumptions allow us to discover novel structure, we also outperform standard deep learning approaches~\cite{pmid28437486}, both in head-to-head performance comparison (mean AUC $\approx 90\%$ vs $85\%$), and the fact that we can produce AUCs for individual locations, have much lower sample complexity, and require only past incidence data. The ability to predict future events based purely on past event streams allows us to probe the underlying social constructs via injecting perturbations in the rates of the different event categories. This is either not feasible, or at least not transparently so,  if a more diverse  set of features are used as inputs; it is hard to see for example how one perturbs street imagery in a principled manner, which has been used in the literature~\cite{pmid28437486} to identify problematic neighborhoods from the presence of wall graffiti. Ability to distill accurate predictions using inputs that are easily and interpretably perturbed \textit{in silico} provides us with a new investigative tool in computational sociology.

Our perturbation analyses reveal that the stress response of the city potentially has  indications of socio-economic bias; wealthier neighborhoods respond appropriately when crime rates are elevated, whereas in disadvantaged neighborhoods predicted arrest rates decrease rapidly. Importantly, we are not simply comparing magnitudes here: the neighborhoods in which  direction of movement for the change in arrest rate  is incorrect tends to be the wealthier ones, away from the inner city. A possible explanation is resource constraints on law enforcement, which combined with biased prioritization to wealthier neighborhoods, leads to reduced enforcement efforts in the rest of the city. This is not entirely surprising, and
reinforces  aspects of the notion of
suburban bias in US cities~\cite{meyer16}.

The suburban bias hypothesis grew out of the  idea of urban bias~\cite{lipton1977poor} put forward in the late 1970s, which posited that urban interests wield political power to bias resource allocation. While imagined to be applicable at the much larger scale of countries and nations,
 the existence of a similar effect in US urban society, where the political power and influence concentrates in the  suburbia instead, has long been suspected and written about~\cite{jackson1987crabgrass,duany2001suburban,logan2002suburban,lazare2001america,young2002inclusion}.
Our analysis now  provides direct validation to that effect, which shows up robustly for all years of analysis going back over one and half decades in Chicago (for which we have data).

Additionally, the structure of the inferred perturbation space suggest how violent and property crime rates interact and co-evolve. The fact that we seem to be able to suppress violent crime by increasing the property crime rate has  deep policy implications: it   suggests that proactive policing that penalizes property crimes more aggressively | and thus ``increases'' the rate of property crime by causing to augment the number of such records in the event log | might indeed have a suppressive effect on  more serious crime. This directly echoes some aspects of Wilson and Kelling's \textit{broken windows policy}~\cite{bw82} model that highlighted  the importance of perceived social disorder in setting up conditions conducive for more serious crime.

However, the efficacy of the broken windows model, with regards to both its theoretical validity and its implementation, has been strongly questioned, including by the  original authors~\cite{kelling96}. One of the most prominent adoptions of the broken windows approach to crime and disorder  occurred in New York City. While  crime rates were significantly reduced post-adoption, there is little consensus on the impact of the specific policies that were
introduced, with the qualitative estimates ranging from large~\cite{bratton2009turnaround} to small but significant~\cite{mess07}, to inconclusive~\cite{ros14}, to nil~\cite{harcourt06}. Recently,  quantitative assessment with $30$ randomized experimental and quasi-experimental tests of disorder policing concluded  that such strategies are associated with an overall statistically significant, modest crime reduction effect~\cite{braga15}. We view our results to be further quantitative evidence for this basic notion underlying the broken-windows model. While we do not claim to  resolve or justify the issues stemming from over-zealous or mis-interpreted and possibly mis-applied zero tolerance policing in lieu of Kelling's original ideas, it does indicate that the different categories of crime have limiting cross-dependencies.

This study also sheds light on the question of choosing the right spatial unit or scale of analysis in urban social modeling and in studies
relating to the ecology of crime~\cite{Boessen2015399,Hipp2007659}.
Our results indicate that different spatial locations have different range of local influence; indeed what is meant by the ``local neighborhood'' is context-dependent.
The importance of multiple spatial scales in urban crime has been long suspected~\cite{wool2002,bursick94}, and has been
explored to be as narrow as face blocks~\cite{hunter72} to city communities to census tracts to tract groups~\cite{miethe1994crime}.
Our results indicate that instead of deciding these spatial scales a priori, a better alternative is that the appropriate neighborhoods of predictive influence be  inferred from data, as shown in Fig.~\ref{fig3}, plates A and B.
These results suggest that spatial influence on crime might have a scale-free organization, where no particular unit of organization emerges as particularly important. This interplay of scales, emerging from diverse mechanisms of social interaction is well known~\cite{sherman89,mears06,Weisburd2014,Quick2019339}; this study underscores the importance of these multi-scale processes to determine the optimal influence neighborhoods that vary across the city. Also, the influence of crime seems to behave as a diffusive process only on average,  hence is not very useful for prediction of individual events at specific locations (See Fig.~\ref{fig3}, plate C), and simply assuming diffusive transport mechanisms are therefore incorrect.

Finally,  the Granger net approach is shown to be
successfully predicting future terror attacks via modeling the  GTD. With regards to the AUC we achieve in this case (approximately $80\%$), it is important to remember that our validation is extrapolative. As pointed out earlier, predicting event/no-event on a randomly selected sample of points within two fixed points of time is generally easier. This interpolative exercise need not model the data as an evolving time series, and generally achieves inflated performance values. Evaluating the predictive performance on events occurring at future time-periods requires the models to actually learn the evolution, and the structure of the historical dependencies. To the best of our knowledge, attempts at such extrapolative event-level prediction of terrorism has not been reported, with the only recent reports solving interpolative  exercises with standard machine learning tools~\cite{hao19}, often without a temporal component~\cite{pmid28591138}, or evaluating performance with one-sided metrics such as precision~\cite{mo17}.

\section*{Conclusion}
This study demonstrates that given enough observations, complex social interactions are surprisingly predictable. This opens the door to new approaches to policing, interventions, and policy design methodologies that has the potential to radically improve societal well-being. At the same time, there is the distinct danger of  misuse via  over-zealous enforcement, and careful consideration and transparency needs to be in place for such technologies to be used to make decisions in  public life.

\section*{Acknowledgments}
Our work greatly benefited from discussion of everyone who participated in our workshop series on crime prediction at the Neubauer Collegium for culture and society~\cite{nrm}, and with those with whom we had extended conversations to ground and refine our modeling approach.

Data was provided by the City of Chicago Data Portal at \href{https://data.cityofchicago.org}{https://data.cityofchicago.org}. The City of Chicago (“City”) voluntarily provides the data on this website as a service to the public.  The City makes no warranty, representation, or guaranty as to the content, accuracy, timeliness, or completeness of any of the data provided at this website (\href{https://www.chicago.gov/city/en/narr/foia/data\_disclaimer.html}{https://www.chicago.gov/city/en/narr/foia/data\_disclaimer.html}), and the authors of this study are solely responsible for the opinions and conclusions expressed in this study.

Data on terror attacks was downloaded from the GTD (\href{https://www.start.umd.edu/data-tools/global-terrorism-database-gtd}{https://www.start.umd.edu/data-tools/global-terrorism-database-gtd}), which  is a database of incidents of terrorism from 1970 -2016. The database is maintained by the National Consortium for the Study of Terrorism and Responses to Terrorism (START) at the University of Maryland, College Park in the United States, and receives funding from a variety of organizations including the US Department of Defense, and the National Science Foundation.

This work is funded in part by the Defense Advanced Research Projects Agency (DARPA) project \#FP070943-01-PR and the Neubauer Collegium for Culture and Society through the Faculty Initiated Research Program 2017. The claims made in this study  do not necessarily reflect the position or the
policy of the sponsors, and no official endorsement should be inferred.

\bibliographystyle{naturemag}
\bibliography{BibLib1,crime,Bibcaus}

\end{document}

%% file: preamble.tex
\usepackage{etex}
\usepackage{amssymb,amsfonts,amsmath,amsthm}
\usepackage{graphicx}
 \usepackage[usenames,x11names, dvipsnames, svgnames]{xcolor}
\usepackage{amsmath,amssymb}
\usepackage{dsfont}
\usepackage{amsfonts}
\usepackage{mathrsfs}
\usepackage{hyperref}
\hypersetup{
    colorlinks=true,
    linkcolor=black,
    citecolor=black,
    filecolor=black,
    urlcolor=DodgerBlue4,
    breaklinks=false,
}
\usepackage{array}
\usepackage{xr}
\usepackage{tikz}
\usepackage{pgfplots}
\usetikzlibrary{shapes,calc,shadows,fadings,arrows,decorations.pathreplacing,automata,positioning}
\usetikzlibrary{external}
\usetikzlibrary{decorations.text}
\tikzexternaldisable
\usepackage{geometry}
\usepackage{rotating}
 \definecolor{nodecol}{RGB}{240,240,220}
 \definecolor{nodeedge}{RGB}{240,240,225}
  \definecolor{edgecol}{RGB}{130,130,130}
    \tikzset{%
fshadow/.style={      preaction={
         fill=black,opacity=.3,
         path fading=circle with fuzzy edge 20 percent,
         transform canvas={xshift=1mm,yshift=-1mm}
       }} 
}
\usetikzlibrary{pgfplots.dateplot}
 \usetikzlibrary{patterns}
\usetikzlibrary{decorations.markings}
\usepackage{fancyhdr}
\usepackage{mathtools}
\usepackage{datetime}
\usepackage{comment}

\newcommand{\cgather}[2][\EQSP]{\begingroup\setlength\abovedisplayskip{#1}\setlength\belowdisplayskip{#1}\begin{gather} #2 \end{gather}\endgroup}

\newif\ifproof
\prooffalse 

\usepackage[linesnumbered,ruled,vlined,noend]{algorithm2e}
\newcommand{\captionN}[1]{\caption{\color{darkgray} \sffamily \fontsize{9}{11}\selectfont #1  }}

\usepackage{txfonts}
\renewcommand{\rmdefault}{phv} 
\renewcommand{\sfdefault}{phv} 
\edef\keptrmdefault{\rmdefault}
\edef\keptsfdefault{\sfdefault}
\edef\keptttdefault{\ttdefault}

\usepackage[OT1]{fontenc}
\usepackage{eulervm} 
\edef\rmdefault{\keptrmdefault}
\edef\sfdefault{\keptsfdefault}
\edef\ttdefault{\keptttdefault}
\tikzexternalenable
\tikzfading[name=fade out,
            inner color=transparent!0,
            outer color=transparent!100]

\tikzset{wiggle/.style={decorate, decoration={random steps, amplitude=10pt}}}
\usetikzlibrary{decorations.pathmorphing}
\pgfdeclaredecoration{Snake}{initial}
{
  \state{initial}[switch if less than=+.625\pgfdecorationsegmentlength to final,
                  width=+.3125\pgfdecorationsegmentlength,
                  next state=down]{
    \pgfpathmoveto{\pgfqpoint{0pt}{\pgfdecorationsegmentamplitude}}
  }
  \state{down}[switch if less than=+.8125\pgfdecorationsegmentlength to end down,
               width=+.5\pgfdecorationsegmentlength,
               next state=up]{
    \pgfpathcosine{\pgfqpoint{.25\pgfdecorationsegmentlength}{-1\pgfdecorationsegmentamplitude}}
    \pgfpathsine{\pgfqpoint{.25\pgfdecorationsegmentlength}{-1\pgfdecorationsegmentamplitude}}
  }
  \state{up}[switch if less than=+.8125\pgfdecorationsegmentlength to end up,
             width=+.5\pgfdecorationsegmentlength,
             next state=down]{
    \pgfpathcosine{\pgfqpoint{.25\pgfdecorationsegmentlength}{\pgfdecorationsegmentamplitude}}
    \pgfpathsine{\pgfqpoint{.25\pgfdecorationsegmentlength}{\pgfdecorationsegmentamplitude}}
  }
  \state{end down}[width=+.3125\pgfdecorationsegmentlength,
                   next state=final]{
     \pgfpathcosine{\pgfqpoint{.15625\pgfdecorationsegmentlength}{-.5\pgfdecorationsegmentamplitude}}
     \pgfpathsine{\pgfqpoint{.15625\pgfdecorationsegmentlength}{-.5\pgfdecorationsegmentamplitude}}
  }
  \state{end up}[width=+.3125\pgfdecorationsegmentlength,
                 next state=final]{
     \pgfpathcosine{\pgfqpoint{.15625\pgfdecorationsegmentlength}{.5\pgfdecorationsegmentamplitude}}
     \pgfpathsine{\pgfqpoint{.15625\pgfdecorationsegmentlength}{.5\pgfdecorationsegmentamplitude}}
  }
  \state{final}{\pgfpathlineto{\pgfpointdecoratedpathlast}}
}
\newcolumntype{L}[1]{>{\rule{0pt}{2ex}\raggedright\let\newline\\\arraybackslash\hspace{0pt}}m{#1}}
\newcolumntype{C}[1]{>{\rule{0pt}{2ex}\centering\let\newline\\\arraybackslash\hspace{0pt}}m{#1}}
\newcolumntype{R}[1]{>{\rule{0pt}{2ex}\raggedleft\let\newline\\\arraybackslash\hspace{0pt}}m{#1}}

\def\DISCLOSURE#1{\def\disclosure{#1}}
\DISCLOSURE{\raisebox{15pt}{$\phantom{XxxX}$This sheet contains proprietary information 
 not to be released to third parties except for the explicit purpose of evaluation.}
}

